# Potential and limits of superlattice multipliers coupled to different input power sources

**A. Apostolakis, M. F. Pereira**[*]

Department of Condensed Matter Theory, Institute of Physics, Academy of Sciences of the Czech Republic, Na Slovance 2, 182 21 Praha 8, Czech Republic

**Abstract**. A theoretical study is presented to assess the performance of semiconductor superlattice multipliers as a function of the currently available input power sources. The prime devices which are considered as input power sources are Impatt diodes, InP Gunn devices, superlattice electron devices and Backward Wave Oscillator sources. These sources have been successfully designed to deliver input radiation frequencies in the range from 0.1 to 0.5 THz. We discuss the harmonic power generation of both odd and even harmonics by implementing an ansatz solution stemmed from a hybrid approach combining nonequilibrium Green's functions and the Boltzmann kinetic equation.

**\***M. F. Pereira**,** E-mail: pereira@fzu.cz

## 1 Introduction

The development of frequency multipliers based on the semiconductor superlattices (SSLs) contribute toward the development of efficient devices which can generate high-frequency radiation at gigahertz (GHz) and terahertz (THz) frequencies.[1-8] Typically, the spectral region below 0.1 THz is covered by devices such as Schottky diode multipliers,[9] InP Gunn sources and oscillators,[10,11] Impatt diodes[11-12] and superlattice electron devices (SLED)[13] that all rely on electron transport. In particular, SLED has attracted much attention because it can operate as an efficient millimeter-wave oscillator utilizing the underlying physical processes of Bloch oscillations and domain formations[3] which are involved in miniband transport delivering high power output in the 240-320 GHz range.[13] On the other hand, devices based on optical transitions[1] are used in order to generate terahertz radiation which is extended beyond the mid-infrared region.[14] Semiconductor Superlattice (SSL) multipliers are in the core of the scientific interest due to their electrooptical properties which can be tuned to cover both the GHz and THz range.[2,3]

Several important questions regarding the design of efficient superlattice multipliers remain or they are only partially addressed. For instance, SSL multipliers have been systematically used in combination with other electron devices which function as input radiation sources.[2,5] Thus, the multiplier devices are subjected to limitations due to the coupling architecture and the waveguide structures responsible for the coupling of the input oscillating field into the SSL sample.[5] Other issues are the physical origin of high harmonic generation (HHG) and the exact physical mechanisms that contribute to the high-frequency (HF) nonlinearity of the superlattice devices. The essential idea in such kind of devices is that a sufficiently strong oscillating field couples energy into the material system, which in turn is converted to radiation of odd or even order harmonics of the oscillating frequency due to SSL nonlinearities.[2,5-8] In particular, the ac-driven electrons in the superlattice can display the relevance of the highly nonparabolic miniband dispersion at the band edges by revealing a regime of optical properties,[15] which would otherwise be hindered by relaxation processes in conventional bulk semiconductors. Similarly, the non-perturbative HHG in bulk solids originates partially from intraband contributions to the nonlinear current due to non-parabolic band dispersions.[16] Another possible mechanism for HHG stems from interband excitations due to induced polarization between valence and conductions bands and enhanced many body effects.[17] Furthermore, THz emission arising from Bloch oscillations in a narrow-miniband SSL in the presence of excitonic effects was discussed in Ref. 18. The theory describing the coherent response of a SSL in combined static and THz along-axis electric fields with the inclusion of excitonic effects was presented in Ref. 19.

At this point, is worth noting that nonlinear optical effects in semiconductors have been discussed thoroughly in the near infrared and visible spectra, using different versions of semiconductor Bloch equations and NEGF methods for both interband[20-29] and intersubband cases[30-37] where the optical response is due to transitions between well-defined subbands in both quantum wells, SSLs and quantum cascade devices. Here in contrast we exploit rather different concept and mechanism, in which the current within a single SSL miniband gives rise to the optical response and nonlinear effects, by means of frequency multiplication in SSLs. We consider the potential performance at room temperature for different high power input radiation sources which have been the subject of recent experimental studies[2,10-13] and stable material systems that their design relies on commercially available engineering components.[38]

## 2 Theoretical Method

The miniband electron transport is described by a hybrid model that combines the Boltzmann equation in a relaxation rate-type approximation with input calculated by the non-equilibrium Green's functions (NEGF) approach.[2-3] This will be the basis for discussing the response of the superlattice subjected to an electric field $E(t) = E_{dc} + E_{ac}\cos(2\pi\nu t)$. Here $\nu$ is the alternating field frequency, $E_{ac}$ is the amplitude of the ac field and $E_{dc}$ is the static field, which corresponds to a constant voltage over the SSL with lattice period $d$. In the case of this oscillating field, $E(t)$, one can determine periodic solutions of the general current response which reads

$$j(t) = j_{dc} + \sum_{l=-\infty}^{\infty} j_l^c \cos(2\pi\nu l t) + j_l^s \sin(2\pi\nu l t),$$

$$j_{dc} = \sum_{p=-\infty}^{\infty} J_p^2(\alpha) Y(U),$$

$$j_l^c = \sum_{p=-\infty}^{\infty} J_p(\alpha)[J_{p+l}(a) + J_{p-l}(a)]\, Y(U),$$

$$j_l^s = \sum_{p=-\infty}^{\infty} J_p(\alpha)[J_{p+l}(a) - J_{p-l}(a)]\, K(U), \qquad (1)$$

where $j_{dc}$ is the dc current, $j_l^c, j_l^s$ are the spectral components and $J_p$ is the Bessel function of the first kind and order $p$.[2,5] The parameter $U = u + ph\nu$, where $u = eE_{dc}d$ represents energy drop per period that electron experiences and $\alpha = eE_{ac}\,d/h\nu$ denotes the control parameter for the conversion of oscillating field into higher harmonics of even and odd order. In Ref. 5, the current response $j(t)$ was calculated using the same formalism [see Eq. (1)] and accurately compared to experimental data for different input frequencies and power delivered only by a BWO. Thus, this approach allows us to investigate in detail the possibility of integrating SSL multipliers with other input sources to achieve realistic GHz-THz devices. The underlying assumption of the relaxation rate approximation[3] provides a compact way of representing functions $Y$ and $K$ as

$$Y(U) = 2j_0 \frac{U/\Gamma}{1+(U/\Gamma)^2}, \quad K(U) = \frac{2j_0}{1+(U/\Gamma)^2}, \qquad (2)$$

where

$$j_0 = e\frac{\Delta d}{2\hbar}\frac{1}{(2\pi)^3}\int dk_z \int d^2k \cos(k_x d)\, n_F(\mathbf{k}, k_z). \qquad (3)$$

Here $\Gamma$ and $j_0$ are the scattering rate and the peak current, respectively, $k_x$ is the projection of the quasimomentum on the $z$ − axis (principal growth direction of the SL), $\mathbf{k} = (k_x, k_y)$ denotes the quasimomentum in the $x − y$ plane and $n_F(\mathbf{k}, \mathbf{k_z})$ is the Fermi distribution. Note that $k_z$ is integrated over the Brillouin zone and the integration limits for $k_y$ and $k_x$ are $\mp\infty$. The power emitted can be determined by the harmonic currents $j_l^c, j_l^s$ which allow the calculation of the averaged Poynting vector.[5,39] Therefore, we obtain that the output power of a tight-binding SL multiplier under the influence electric field is

$$P_l(\alpha, \nu) = \frac{A\,\mu_0\,c\,L^2}{8\,n_r} I_l(\alpha, \nu), \qquad (4)$$

where $I_l^2(\nu) = (j_l^c)^2 + (j_l^s)^2$, $n_r$ is the refractive index of the SL sample, A is the contact area of the device converting current density into current, $\mu_0$ is the permeability in the free space, $c$ is the speed of light in the free space and $L$ is the effective path through the crystal. In this work, to determine the generated power we chose the parameters of a realistic GaAs/AlGaAs SSL,[5] namely, miniband witdh $\Delta = 140$ meV, period $d$=6.23 nm, electron density $N_0 = 1.5 \times 10^{18}$ cm$^{-3}$, refractive index $n_r = \sqrt{13}$ (GaAs) and the relaxation time $\tau=\hbar/\Gamma$=31 fs. The parameters extracted directly from NEGF calculations[2] for a symmetric SL structure and used in the relaxation rate-type approximation for the Boltzmann equation were: $\Gamma$ =21 meV and $j_0$=2.14$\times 10^9$ A/m$^2$.

A previous study has shown that frequency multiplication effects in a voltage-biased superlattice are pronounced when Bragg oscillating electrons directly interact with the input electromagnetic field.[8] Enhancement the generated power was attributed to the frequency modulation of Bloch oscillations which are triggered under Negative Differential Conductivity (NDC) conditions. Let us underline here that in the present study we consider a plain harmonic field ($E_{ac} \cos 2\pi\nu t$) and thus Bragg reflections from the minizone boundaries are not related to a specific oscillation period ($\nu_B = eE_{dc}d/h$) as in the case of the static field. On the contrary these reflections result in frequency modulation of the electron oscillations[40] during a time-period ($T = 1/\nu$) determined strictly by the frequency of the oscillating field. There question that arises is what is the exact condition for the development of Bloch oscillations in a harmonic field. In this case the onset of these phase-modulated Bloch oscillations is determined by the criterion $\alpha > \alpha_c$, where $\alpha_c = U_c/h\nu$ and $U_c = \Gamma$. Therefore, a strong ac-field brings the superlattice to an active state, i.e. the NDC region of the current voltage characteristic. Note that this criterion ($\alpha > \alpha_c$) NDC works most efficiently in the presence of low-frequency electric field $\frac{h\nu}{\Gamma} \ll 1$ and under the condition $\frac{\left(\frac{\hbar}{\Gamma}\right)dE(t)}{dt} \ll U_c$,[41] so that the current density follows almost adiabatically the temporal value of the electric field according to Eq. (2). For the input power sources considered in the present work, the condition $\frac{h\nu}{\Gamma} < 1$ is predominately satisfied. For instance, the InP Gunn devices can generate an oscillating field with maximum frequency ($\nu$=479 GHz, see Table 1) which satisfies the aforementioned condition ($\frac{h\nu}{\Gamma} \sim 0.0943$).

The SSL NDC state might be accompanied by additional nonlinearities in the form of high-field domains[7] similar to NDC states of other bulk materials. [42] The generation of harmonics in SSLs due to the periodic formation and annihilation of electric domains has been studied both experimentally[7,13] and theoretically[7]. The basic principle behind these effects is that during each half-period the field, a domain is created when the absolute value of the instant field strength exceeds $E_c = \Gamma/(ed)$ and suppressed when becomes smaller than $E_c$.[7] The successive creation and annihilation of domains contributes to a high-frequency current which is the source of the high-harmonics radiation. Furthermore, the time characterizing the growth and collapse of space-charge domains in superlattices might not only be determined by the intraminiband relaxation rate $\Gamma$, but also by the dielectric relaxation rate $\Gamma_{\text{diel}} = 4\pi\sigma_{dc}/\epsilon\epsilon_0$ where $\epsilon$ is the relative permittivity and $\epsilon_0$ is the permittivity of the vacuum and $\sigma_{dc}(E_{dc}) = \frac{dj_{dc}}{dE_{dc}} = (\sigma_0/2j_0)\frac{\partial Y(u)}{\partial u}$ is the dc differential conductivity.[43] Using a similar treatment as in the natural bulk semiconductors,[44] one can assume that the characteristic time domain formation $\Gamma_{\text{dom}}$ is the smaller rate between $\Gamma_{\text{diel}}$ and $\Gamma$, i.e. $\Gamma_{\text{dom}} = \min\{\Gamma, \Gamma_{\text{diel}}\}$.[43] Taking into consideration Eq. (2), the dielectric relaxation rate takes the form $\Gamma_{\text{diel}} = \frac{\hbar\omega_{pl}^2}{\Gamma}\left(1 - \left(\frac{u}{\Gamma}\right)^2\right)\left[1 + \left(\frac{u}{\Gamma}\right)^2\right]^{-2}$ where $\omega_{pl} = \left[\frac{2j_0 ed}{(\hbar\epsilon\epsilon_0)}\right]^{1/2}$ is the miniband plasma frequency.[45] The dc-conductivity takes its manimum value $\min[\sigma_{dc}(u)] = -\frac{\sigma_0}{8}$ at $\frac{u}{\Gamma} = \sqrt{3}$, where $\sigma_0 = 2j_0 ed/\Gamma$ is the Drude conductivity and therefore the dielectric relaxation rate is $\Gamma_{\text{diel}} \sim 4.5$ meV $> \Gamma$ for the SSL structure considered here. By choosing a heavily doped SL with a very wide miniband, similar to the parameters used in our calculations, results in an increase of $\Gamma_{\text{diel}}$ and plasma effects playing a minimal role in exciting the electrons in the NDC state.

We conclude this section by underlying that in the present study we assume that the distribution of electrons is approximately homogeneous and therefore the local electric field-current density relation is identical with the global voltage-current (VI) characteristic [see Eqs. (1)-(3)]. Thus, the radiation of high-frequency fields at odd harmonics of the ac-field stems from the nonlinearity of the voltage-current characteristic and the phase-modulated Bloch oscillations.

## 3 Numerical results and discussion

After obtaining a satisfactory model for analysis of the general current response, the next logical step is to estimate the response of the SSL multiplier for different power inputs. Before we proceed it is worth noting that here we use direct applications of our equations for optimized solutions, but in future research we plan to use more advanced optimization methods[47-49] for integrated sources and multipliers. Table 1 shows the magnitude of the output power of sources with optimized frequency which allows them to be used together with multipliers to reach frequencies in the far terahertz spectral range. Note that all simulations presented here are without a static bias, i.e., $u = eE_{dc}d = 0$.

Table 1 Summary of the parameters for the GHz-THz input sources for the excitation SSL multipliers.

| Types of power input sources | Spectral frequencies | $P_{in}$ | References |
|---|---|---|---|
| BWO | 130, 141, 150, 160 GHz | 61.6, 47, 45.2 42.2 µW | M. F. Pereira et. al.[2,5] |
| SLED | 127.1, 145.3, 156.5 GHz<br>249.6, 317.4 GHz | 14, 4.2, 1.7 mW<br>0.92 mW, 77 µW | H. Eisele et. al.[10,11] |
| InP Gunn devices | 193, 412, 479 GHz | 34 mW, 330 µW, 86 µW | H. Eisele et. al.[13] |
| Impatt diodes (TeraSense) | 100, 140, 290 GHz | 80, 30, 10 mW | see Ref. 38 |

The list of sources given in Table 1 combines research devices which have attracted recent attention[2,10-13] with commercially available generators.[38] We note, however, that one of the possible limitations to achieving high power in SSL structures is associated with the whole coupling setup between the multipliers and the GHz input sources. Therefore, possible losses arise due to impedance mismatch between the superlattice element and the external waveguide system.[5,8] Figure 1 outlines the calculations for the third harmonics which can be extracted from a SSL multiplier after the excitation with input power delivered by the input sources of Table I. The highest third-harmonic frequency is 1437 GHz with an output power of approximately 1.45 µW. This harmonic power output can be generated from a superlattice in combination with a InP Gunn device which operates as an input source.[10] An improved performance of 5.7 µw at 1236 GHz is expected for a SSL multiplier, assuming an InP-based input source with 330 µW.[11] The output power of 6.8 µW at 749 GHz constitutes the strongest response considering the integration with a SLED oscillator operating in the fundamental mode.[13] The SSL multipliers of Fig. 1 were also numerically investigated for their power output which can potentially be obtained for the 5-th harmonic component. Again, the harmonic power extraction [see Eq. (4)] from a superlattice multiplier under the influence of an oscillating field with frequency 156.5 GHz which is generated by a SLED[13] source exceeds the ones produced by the other input sources [see Fig.2]. Note that output powers tend to decrease significantly when a SSL multiplier is pumped with a higher RF output power from the SLED oscillator. The powers which may be detected at the 7th-harmonic frequency drop significantly compared with the lower-order responses as shown in Fig. 3. However, with an Impatt input source[38] operating at 290 GHz at 10 mW, the SSL frequency multiplier can generate a 7th-harmonic with magnitude similar to the 5-th or 3-rd harmonics. In addition, this reported output power becomes comparable with the 7-th harmonic of a SL multiplier integrated with SLED oscillators. The latter frequency multiplier system can exceed the output power of 1.312 μW at 785.2 GHz.

Note that a SSL characterized by perfectly antisymmetric current flow, when irradiated by an ac-field can only spontaneously generate odd harmonics. However, as demonstrated in the Appendix the comparison of SSL multiplier's performance for different input sources is feasible for even harmonics by considering imperfections in the structure, which lead to asymmetric current flow.

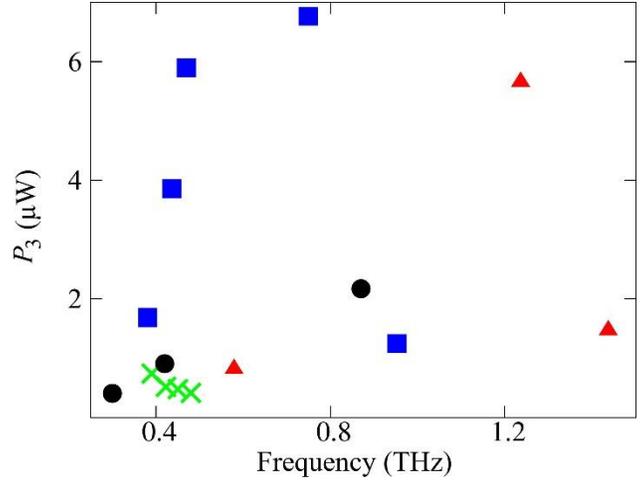

**Fig. 1** Third harmonic output powers from SSL multipliers for different input sources ● (Impatt diodes), ■ (SLED devices), ▲ (InP Gunn devices) and ✕ (BWO sources) over the frequency range 300 GHz-1600 GHz. The input field power in each case corresponds to the power generated by the devices given in Table 1.

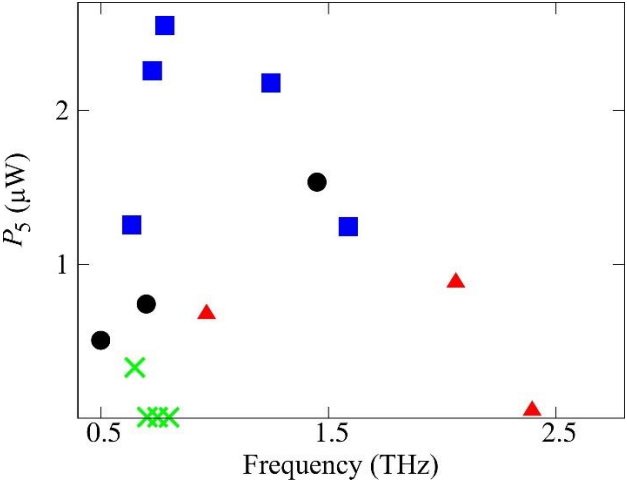

**Fig. 2** Fifth harmonic output powers from SSL multipliers for different input sources ● (Impatt diodes), ■ (SLED devices), ▲ (InP Gunn devices) and ✕ (BWO sources) over the frequency range 500 GHz-2400 GHz. The input field power in each case corresponds to the power generated by the devices given in Table 1.

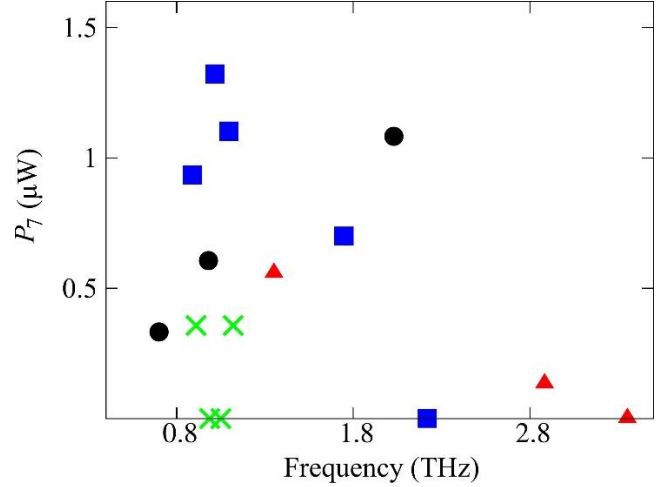

**Fig. 3** Seventh harmonic output powers from SSL multipliers for different input sources ● (Impatt diodes), ■ (SLED devices), ▲ (InP Gunn devices) and ✕ (BWO sources) over the frequency range 700 GHz-3000 GHz. The input field power in each case corresponds to the power generated by the devices given in Table 1.

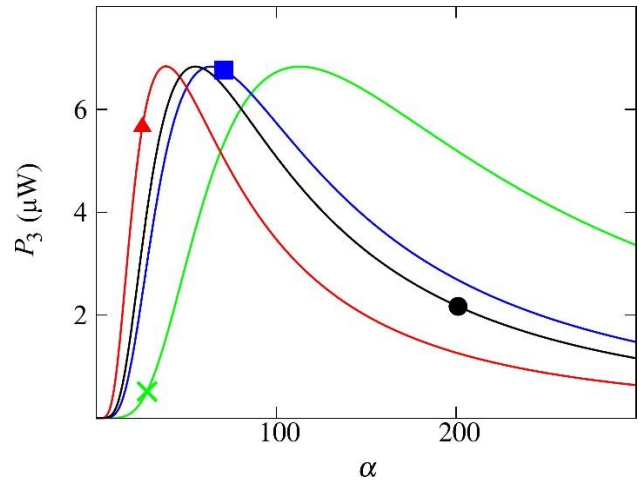

**Fig. 4** Third harmonic power output as a function of $\alpha = eE_{ac}\, d/h\nu$ for SSL multipliers which convert input fields with different oscillating frequencies. The input frequencies $\nu =$ 130, 249.6, 290, 412 GHz correspond to the green, blue, black and red curves, following the color convention used for each device. The symbols denote the maximum output depicted in Fig. 1 for each input source separately, namely ● (Impatt diodes), ■ (SLED devices), ▲ (InP Gunn devices) and ✕ (BWO sources).

Figures 5 and 6 complement Figs. 2 and 3 by demonstrating the conversion of the input oscillating field as a function of the parameter $\alpha$ to the 5-th and 7-th harmonic respectively. Evidently the maximum power output of the 5-th and 7-th harmonic is significantly smaller than the magnitude of the 3-rd order harmonic [see Fig. 3]. In addition, we expect that the maximum harmonic signals (5-th and 7-th harmonics) require remarkably larger $\alpha$ in comparison to the 3-rd harmonic and therefore input devices delivering higher input power.

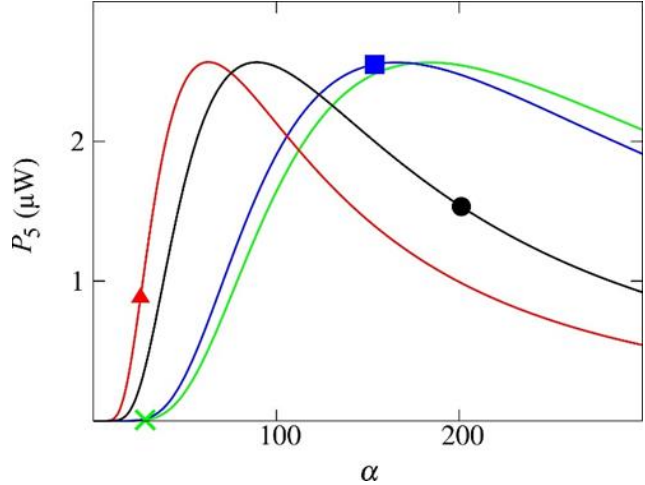

**Fig. 5** Fifth harmonic power output as a function of $\alpha = eE_{ac}\, d/h\nu$ for SSL multipliers which convert input fields with different oscillating frequencies. The input frequencies $\nu =$130, 156.5, 290, 412 GHz, correspond to the green, blue, black and red curves, following the color convention used for each device. The symbols denote the maximum output depicted in Fig. 2 for each input source separately, namely ● (Impatt diodes), ■ (SLED devices), ▲ (InP Gunn devices) and ✕ (BWO sources).

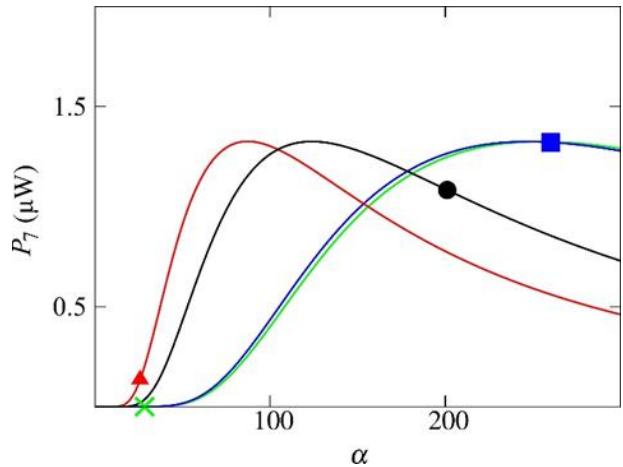

**Fig. 6** Seventh harmonic power output as a function of $\alpha = eE_{ac}\, d/h\nu$ for SSL multipliers which convert input fields with different oscillating frequencies. The input frequencies $\nu =$140, 145.3, 193, 290 GHz correspond to green, blue, black and red curves, following the color convention used for each device. The symbols denote the maximum output depicted in Fig. 3 for each input source separately, namely ● (Impatt diodes), ■ (SLED devices), ▲ (InP Gunn devices) and ✕ (BWO sources).

In Figs. 5 and 6 we have added the calculations of the largest harmonic outputs for the input power sources summarized in Table 1. For instance, Fig. 5 shows that the maximum output of the 5th-harmonic at 650 GHz is not only achievable for a SLED input source but also if a BWO could provide certain power ( $\sim 4.2$ mW ) to the SSL multiplier. This further confirms the potential of integrating SLED devices with SSL multipliers. Our study is limited by the available working frequency range of the power sources. Further performance improvements are anticipated if the input sources such as Impatt diodes[11-12] or InP Gunn devices[11,13] are designed to generate an optimal output rf power. Note that, by increasing $\alpha$, the output power increases approximately up to 7 µW just after $\alpha$ surpasses $\alpha_c$. We can observe that the power output of a SL multiplier excited by a SLED[13] oscillator operating at frequency 249.6 GHz is very close to the maximum possible value which can obtained for the third-harmonic radiation as depicted in Fig 4. Therefore, in order to benefit from the maximum harmonic response of a superlattice multiplier, the input sources should deliver a power which follows the behavior of $P_l$ $(\alpha, \nu)$. Otherwise, a larger input power could result in a significantly reduced performance of the frequency multiplier even for an oscillating field with similar frequency.

The performance depends on the input power (measured by ) in a complex way, BWO's are useful to use in fundamental studies due to their tunability[2,5], but they are large and expensive. Furthermore, output is clearly maximized by SLEDs, InP Gunn devices and Impatt diodes, which are compact sources that can be more easily combined with SSL multipliers to deliver optimized room temperature sources for the 0.3 to 2.0 THz range.

## 4 Conclusion

In conclusion, the theoretical results reported in this paper confirm the unique potential of SL multipliers which enables up-conversion to sub-THz and THz frequency ranges. Enhanced performance is expected when the output radiation generated by the input source is efficiently coupled into the SL multiplier by reducing losses in waveguide structures. The combination of SLEDS[13] and superlattice multipliers show advantages in the efficiency of harmonic generation providing measurable power up to 7 μW at around 1 THz, operating at room temperature.

**Appendix: Even-order responses of the SSL multipliers**

In the main part of the text (see Sec. 3) we discussed the odd-order responses which are generated after the superlattice multiplier gets excited by an input field having different magnitudes and frequencies. The parameters of the oscillating field were chosen in such a way to comply with the available input power sources (see Table 1). Recently, though, it was shown that spontaneous frequency multiplication effects are feasible for even harmonics in a SSL multiplier.[2] Conventionally, generation of even harmonics can take place only in a biased SSL or due to parametric effects.[3,49] However, by considering the influence of the asymmetry in current flow induced by interface roughness scattering, even harmonic generation can emerge.[2,5,39] The *V-I* characteristics described by Eq. (1) then should be modified according to

$$j_0 = \begin{cases} j_0^-, & U < 0 \\ j_0^+, & U \geq 0 \end{cases}, \qquad \Gamma = \begin{cases} \Gamma^-, & U < 0 \\ \Gamma^+, & U \geq 0. \end{cases}$$

This ansatz solution describes accurately the asymmetry in current-voltage curve by associating the parameters $j_0^+$, $j_0^-$ (peak current density) with the critical energies $U_c^+ = \Gamma^+$ and $U_c^+ = \Gamma^+$ respectively. The above parameters were extracted by exact NEGF calculations and they are in reasonable agreement with experimental measurements.[2] To estimate the power output of even harmonics we use the parameters $\Gamma^+$, $\Gamma^- =21, 20$ meV and $j_0^+$, $j_0^-=2.14, 1.94\times 10^9$ A/m$^2$ as in Refs. 2, 5. Figure 7 depicts estimations for the second and fourth harmonics which can be extracted from a SSL multiplier after the illumination. We can see that second harmonic responses behave significantly different than the odd responses we discussed in Sec. 3. Here the harmonic output from a superlattice in combination with BWO devices[5] exceeds systematically the other coupling setups for discussions of lower-frequency modes. In particular, we expect a performance of 16 nW at 260 GHz, by considering a BWO input source with 61 μW power input. At the same time, it should be underscored that a high-frequency second harmonic with enhanced power output (~ 22 nW) can be generated by utilizing a SLED input source[13] operating at 317 GHz. Now considering the quadruple effects, we note the output power of 4.15 nW at 824 GHz, given an InP Gunn device[10] which couples an input oscillating field within the superlattice with frequency 412 GHz and corresponding input power 330 mW. The conversion efficiency of even harmonics for an unbiased SL multiplier is far weaker from the performance of other available material systems.[9] Nevertheless, as has been discussed recently,[39] the optimization of the asymmetric effects in a heterostructure semiconductor, i.e. a systematic interface roughness design, can lead to significant enhancement of even harmonic output power.

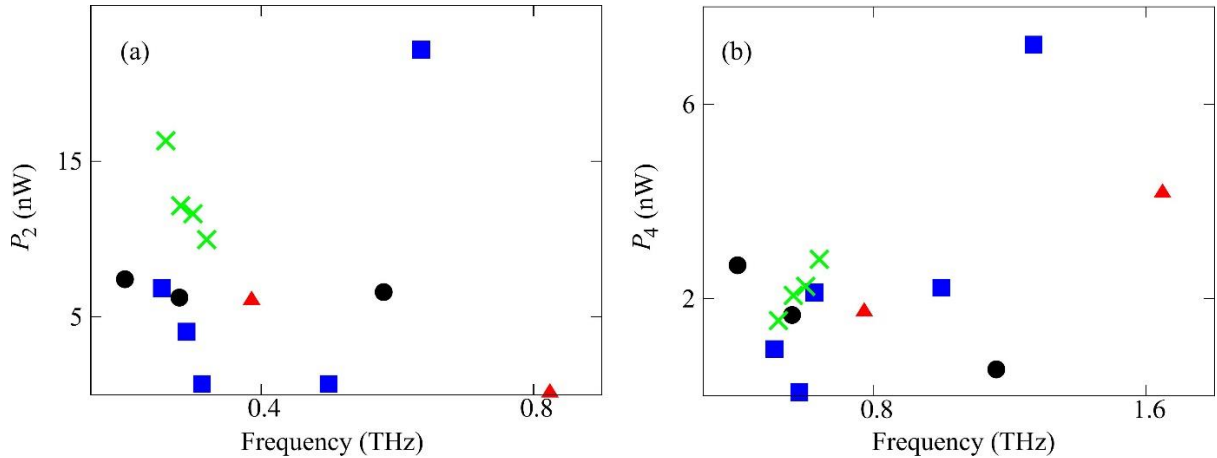

**Fig. 7** (a) Second harmonic output powers from SSL multipliers for different input over the frequency range 200 GHz-900 GHz. (b) Fourth harmonic output powers from SSL multipliers for different input sources over the frequency range 400 GHz-1700 GHz. The input field power in each case corresponds to the power generated by the devices given in Table 1. The symbol convention in both plots is: ● (Impatt diodes), ■ (SLED devices), ▲ (InP Gunn devices) and ✖ (BWO sources).

*Acknowledgments*

The authors acknowledge support by the Czech Science Foundation (GAČR) through grant No. 19-03765 and the EU H2020-Europe's resilience to crises and disasters program [H2020 - grant agreement n˚832876].

*References*

1. M. Razeghi, Q. Y. Lu, N. Bandyopadhyay, W. Zhou, D. Heydari, Y. Bai and S. Slivken, "Quantum cascade lasers: from tool to product," Opt. Express 23, 8462-8475 (2015)
2. M. F. Pereira, D. Winge, A. Wacker, J. P. Zubelli, A. S. Rodrigues, V. Anfertev and V. Vaks, "Theory and Measurements of Harmonic Generation in Semiconductor Superlattices with Applications in the 100 GHz to 1 THz Range," Phys. Rev. B 96, 045306 (2017).
3. A. Wacker, "Semiconductor superlattices: a model system for nonlinear transport," Physics Reports 357, 1-111 (2002).
4. M. F. Pereira, "TERA-MIR Radiation: Materials, Generation, Detection and Applications II," Opt. Quant. Electron. 47, 815-820 (2015).
5. M. F. Pereira, V. Anfertev, J. P. Zubelli and V. Vaks, "THz Generation by GHz Multiplication in Superlattices," J. of Nanophotonics, 11, 046022 (2017).
6. A. Apostolakis, M. F. Pereira, "Numerical studies of superlattice multipliers performance," Quantum Sensing and Nano Electronics and Photonics XVI, 109262G (2019).
7. F. Klappenberger, K. F. Renk, P. Renk, B. Rieder, Yu. I. Koshurinov, D. G. Pavel'ev, V. Ustinov, A. Zhukov, N. Maleev and A. Vasilev, "Semiconductor-superlattice frequency multiplier for generation of submillimeter waves," Appl. Phys. Lett. 84, 3924 (2004).
8. S. Winnerl, E. Schomburg, S. Brandl, O. Kus, K. F. Renk, M. Wanke, S. Allen, A. Ignatov, V. Ustinov, A. Zhukov and P. Kop'ev, "Frequency doubling and tripling of terahertz radiation in a GaAs/AlAs superlattice due to frequency modulation of Bloch oscillations," Appl. Phys. Lett. 77, 1762 (2000).
9. F. Maiwald, F. Lewen, B. Vowinkel, W. Jabs, D. G Pavel'ev, M. Winnerwisser, G. Winnerwisser, "Planar Schottky Diode Frequency Multiplier for Molecular Spectroscopy up to 1.3 THz," IEEE Microwave Guided Wave Lett., 9, 198-200 (1999).

10. H. Eisele, "480 GHz oscillator with an InP Gunn device," Electron. Lett. 46, 422-423 (2010).
11. H. Eisele, "State of the art and future of electronic sources at terahertz frequencies," Electron. Lett. 46, 8-11 (2010).
12. M. Ino, T. Ishibashi and M. Ohmori, "C.W. oscillation with p+-p-n+ silicon IMPATT diodes in 200 GHz and 300 GHz bands", Electron. Lett. 12 (2010).
13. H. Eisele, L. Li and E. H. Linfield, "High-performance GaAs/AlAs superlattice electronic devices in oscillators at frequencies 100–320 GHz," Appl. Phys. Lett. 112, 172103 (2018).
14. M. Tonouchi, "Cutting-edge terahertz technology," Nat. Photonics 1, 97 (2007).
15. Yu. A. Romanov, "Nonlinear effects in periodic semiconductor structures (Frequency multiplication due to nonparabolicity of dispersion law in semiconductor structure)", Opt. Spektrosk. 33, 917 (1972).
16. S. Ghimire, A. D. DiChiara, E. Sistrunk, P. Agostini, L. F. DiMauro, and D. A. Reis, "Observation of high-order harmonic generation in a bulk crystal," Nat. Phys. 7, 138 (2011).
17. O. Schubert, M. Hohenleutner, F. Langer, B. Urbanek, C. Lange, U. Huttner, D. Golde, T. Meier, M. Kira, S. W. Koch, and R. Huber, "Sub-cycle control of terahertz high-harmonic generation by dynamical Bloch oscillations," Nature Photon. 8, 119 (2014).
18. H. T. Duc, T. Meier, and S. W. Koch, "Coherent Electric-Field Effects in Semiconductors," Phys. Rev. Lett. 95, 086606.
19. M. M. Dignam, "Excitonic Bloch oscillations in a terahertz field," Phys. Rev. B, 59 (1999).
20. M. F. Pereira, S. W. Koch and W. W. Chow, "Many-body effects in the gain spectra quantum-wells," Appl. Phys. Lett. 59, 2941-2943 (1991).
21. C. I. Oriaku and M. F. Pereira, "Analytical solutions for semiconductor luminescence including Coulomb with applications to dilute bismides," J. Opt. Soc. Am. B, 34, 321–328 (2017).
22. M. F. Pereira, "Analytical Expressions for Numerical Characterization of Semiconductors per Comparison with Luminescence," Materials 11, 1-15 (2018).
23. H. Grempel, A. Diesel, W. Ebeling, J. Gutowski, K. Schüll, B. Jobst, D. Hommel, M. F. Pereira and K. Henneberger, "High-density effects, stimulated emission, and electrooptical properties of ZnCdSe/ZnSe single quantum wells and laser diodes," Phys. Status Solidi B 194, 199 (1996).
24. M. F. Pereira, R. Binder and S. W. Koch, "Theory of nonlinear optical absorption in coupled-band quantum wells with many-body effects," Appl. Phys. Lett. 64, 279-281 (1991).
25. M. F. Pereira and K. Henneberger, "Microscopic Theory for the Optical Properties of Coulomb-Correlated Semiconductors," Phys. Status Solidi B 206, 477-491 (1998).
26. W. W. Chow, M. F. Pereira and S. W. Koch (1992), "Many-body treatment on the modulation response in a strained quantum well semiconductor laser medium," Appl. Phys. Lett. 61, 758 (1992).

27. R. Jin, K. Okada, G. Khitrova, H. M. Gibbs, M. F. Pereira and S. W. Koch and N. Peyghambarian, "Optical nonlinearities in strained-layer InGaAs/GaAs multiple quantum wells," Appl. Phys. Lett. 61, 1745 (1992).
28. M. F. Pereira and K. Henneberger, "Gain mechanisms and lasing in II-VI compounds," Phys. Status Solidi B 202, 751-762 (1997).
29. M. F. Pereira, "Analytical expressions for numerical characterization of semiconductors per comparison with luminescence," Materials 11, 2 (2018).
30. M. F. Pereira, R. Nelander, A. Wacker, D. G. Revin, M. R. Soulby, L. R. Wilson, J. W. Cockburn, A. B. Krysa, J. S. Roberts and R. J. Airey, "Characterization of intersubband devices combining a nonequilibrium many body theory with transmission with transmission spectroscopy experiments," Sci. Mater. Electron. 18, 689 (2007).
31. M. F. Pereira and S. Tomić, "Intersubband gain without global inversion through dilute nitride band engineering," Appl. Phys. Lett. 98, 061101 (2011).
32. M. F. Pereira and I. A. Faragai, "Coupling of THz radiation with intervalence band transitions in microcavities," Opt. Express 22, 3439 (2014).
33. M. F. Pereira, "Microscopic approach for intersubband-based thermophotovoltaic structures in the terahertz and mid-infrared," JOSA B 28, 2014-2017 (2011).
34. M. F. Pereira and T. Schmielau "Momentum dependent scattering matrix elements in quantum cascade laser transport. Microelectron," J. 40, 869-871 (2009).
35. R. Nelander R, A. Wacker, M. F. Pereira, D. G. Revin, M. R. Soulby, L. R. Wilson, J. W. Cockburn, A. B. Krysa, J. S. Roberts and R. J. Airey, "Fingerprints of spatial charge transfer in quantum cascade lasers," Appl. Phys. 102, 11314 (2007).
36. M. F. Pereira, "The linewidth enhancement factor of intersubband lasers: from a two-level limit to gain without inversion conditions," Appl. Phys. Lett. 109, 222102 (2016).
37. T. Schmielau and M. F. Pereira, "Nonequilibrium many body theory for quantum transport in terahertz quantum cascade lasers," Appl. Phys. Lett. 95, 231111 (2009).
38. *"Terahertz sources," terasense.com*, http://terasense.com/products/terahertz-sources/ (accessed 25 April 2019).
39. A. Apostolakis and M. F. Pereira, "Controlling the harmonic conversion efficiency in semiconductor superlattices by interface roughness design," AIP Advances 9, 015022 (2019).
40. The terminology employed in previous works[41] to described this type of oscillations was Bloch oscillations in a harmonic field (BOHF).
41. Yu. A. Romanov, Yu.Yu Romanova, "Bloch oscillations in superlattices: The problem of a terahertz oscillator," Semiconductors, 39 (2005).
42. H. Kroemer, "Theory of the Gunn effect," Proc. IEEE, 52, 1736 (1964).
43. F. Klappenberger, K. N. Alekseev, K. F. Renk, R. Scheuerer, E. Schomburg, S. J. Allen, G. R. Ramian, J. S. S. Scott, A. Kovsh, V. Ustinov, and A. Zhukov, "Ultrafast creation and annihilation of space-charge domains in a semiconductor superlattice observed by use of Terahertz fields," Eur. Phys. J. B 39, 483 (2004).

44. B.R. Pamplin, "Negative differential conductivity effects in semiconductors," Contemp. Phys.11, 1 (1970).
45. T. Hyart, N. V. Alexeeva, J. Mattas, and K. N. Alekseev, "Terahertz Bloch Oscillator with a Modulated Bias," Phys. Rev. Lett. 102, 140405 (2009).
46. J. Radovanović, V. Milanović, Z. Ikonić and D. Indjin, "Application of the genetic algorithm to the optimized design of semimagnetic semiconductor-based spin-filters," Journal of Physics D: Applied Physics 40, 5066 (2007).
47. J. Radovanović, V. Milanović, Z. Ikonić and D. Indjin, "Optimization of spin-filtering properties in diluted magnetic semiconductor heterostructures," Journal of applied physics 99, 073905 (2006).
48. J. Radovanović, V. Milanović, Z. Ikonić, D. Indjin, "Quantum well shape optimization of continuously graded $Al_xGa_{1-x}N$ structures by combined supersymmetric and coordinate transform methods," Physical Review B 69, 115311 (2004).
49. T. Hyart, A. V. Shorkhov and K. N. Alekseev, "Theory of Parametric Amplification in Superlattices," Phys. Rev. Lett. 98, 220404 (2007).

**A. Apostolakis** received his Ph.D. from the Department of Physics, School of Science, Loughborough University, Loughborough, UK in 2017. He is currently a Post-doctoral researcher at Department of Condensed Matter Theory, Institute of Physics, Academy of Science of Czech Republic, Prague. His main research interests include THz physics, electron transport, semiconductor heterostructures and opto-electronic devices.

**M. F. Pereira** is a Professor and Head of Department of Theory of Condensed Matter, Institute of Physics, Academy of Sciences of Czech Republic. His main research interest is on nonequilibrium many-body theory applied to quantum and nonlinear optics and transport in semiconductor materials and has coauthored over 120 publications. He received his Ph.D. at the Optical Sciences Center, University of Arizona, and is a fellow of SPIE.